\documentclass[sigconf]{acmart}
\AtBeginDocument{%
  }

\setcopyright{acmlicensed}
\copyrightyear{2026}
\acmYear{2026}
\setcopyright{cc}
\setcctype{by}
\acmConference[AST '26]{7th ACM/IEEE International Conference on Automation of Software Test (AST 2026)}{April 13--14, 2026}{Rio de Janeiro, Brazil}
\acmBooktitle{7th ACM/IEEE International Conference on Automation of Software Test (AST 2026) (AST '26), April 13--14, 2026, Rio de Janeiro, Brazil}
\acmPrice{}
\acmDOI{10.1145/3793654.3793756}
\acmISBN{979-8-4007-2476-3/2026/04}

\usepackage{soul}
\usepackage{subcaption}
\usepackage{multirow}
\usepackage{siunitx}
\usepackage[ruled,vlined]{algorithm2e}
\usepackage[inline]{enumitem}

\begin{document}

\title{LLAMAFUZZ: Large Language Model Enhanced Greybox Fuzzing}

\author{Hongxiang Zhang}
\affiliation{%
  \institution{University of California, Davis}
  \city{Davis}
  \state{California}
  \country{USA}
}
\email{hxxzhang@ucdavis.edu}

\author{Yuyang Rong}
\affiliation{%
  \institution{University of California, Davis}
  \city{Davis}
  \state{California}
  \country{USA}
}
\email{PtrRong@ucdavis.edu}

\author{Yifeng He}
\affiliation{%
  \institution{University of California, Davis}
  \city{Davis}
  \state{California}
  \country{USA}
}
\email{yfhe@ucdavis.edu}

\author{Hao Chen}
\affiliation{%
  \institution{University of Hong Kong}
  \city{Hong Kong}
  \country{China}
}
\email{chenho@hku.hk}
\renewcommand{\shortauthors}{Hongxiang et al.}

\definecolor{lightblue}{RGB}{200, 220, 255}
\definecolor{lightgreen}{RGB}{210, 240, 210}
\definecolor{lightgray}{RGB}{220, 220, 220}
\definecolor{lightred}{rgb}{1.0, 0.8, 0.8}
\begin{abstract}
Greybox fuzzing has achieved success in revealing bugs and vulnerabilities in programs. However, bit-level randomized mutation strategies have limited the fuzzer's performance on structured data. Specialized fuzzers can handle specific structured data, but require additional efforts in grammar and suffer from low throughput.
In this paper, we explore the potential of utilizing Large Language Models~(LLMs) to enhance greybox fuzzing for structured data. We utilize the pre-trained knowledge of LLM about data conversion and format to generate new valid inputs. 
We further enhance the LLM on structured formats and mutation strategies by fine-tuning with paired mutation seeds.
Our LLM-enhanced fuzzer, {LLAMAFUZZ}, integrates the power of LLM to understand and mutate structured data to fuzzing. 
Our experiments show that {LLAMAFUZZ} outperformed the state-of-the-art methods on all benchmarks, demonstrating its effectiveness in various scenarios.
\end{abstract}



\received{24 October 2025}
\received[accepted]{5 January 2026}

\maketitle

\section{Introduction}
Fuzz testing, also known as fuzzing, is an automated software testing technique that generates test seeds to discover vulnerabilities in the target software applications. In recent years, greybox fuzzing has drawn much attention because of its effectiveness in discovering new vulnerabilities in many programs. As software systems continue to grow in complexity and evolve at an accelerated pace, the need for adapted test inputs has become increasingly important. While bit-level randomized mutation~\cite{afl,fioraldi2020afl++,honggfuzz,lyu2019mopt,lemieux2018fairfuzz} has achieved a lot, they reached a bottleneck in which traditional greybox fuzzers struggle to effectively generate structured data.

General-purpose greybox fuzzers employ high-throughput bit-level mutation. AFL++~\cite{fioraldi2020afl++}, one of the state-of-the-art greybox fuzzers, combines multiple mutation strategies and scheduling strategies, leading fuzzing to a new level. 
However, when fuzzing applications that require structured input, blind random bit-level mutation can be problematic. Such mutations often disrupt the integrity of data formats, resulting in inefficient exploration of the input space. As a result, converging to a high and stable coverage and reaching bugs is time-consuming. To accelerate this process, honggfuzz~\cite{honggfuzz} shares the file corpus across multi-process and multi-thread execution to boost throughput. However, the effectiveness of merely increasing throughput and adding new random mutations is limited,
since the bottleneck is caused by the complex constraints when handling structured seeds. AFL++ and honggfuzz require excessive attempts to generate valid structured inputs. 
Therefore, structural awareness of the test seeds is the key to success in fuzzing such software.
To generate structured binary seeds, specialized fuzzers use predefined grammars. Gramatron~\cite{srivastava2021gramatron} restructures the grammar to enable unbiased sampling from the input state space and permits more aggressive mutation operations. Moreover, it combines search-based testing to co-evolve the test case generation. However, it requires additional specifications in pre-defined Chomsky Normal Form~\cite{chomsky1959certain} and Greibach Normal Form~\cite{greibach1965new} to construct grammar automata. 
NAUTILUS~\cite{aschermann2019nautilus} employs grammar-aware mutation operators to probe deep program paths and reveal complex bugs, while GRIMOIRE~\cite{srivastava2021gramatron} synthesizes input structures to discover bugs in grammar-based formats.

\begin{figure}[ht]
     \centering
       \includegraphics[width=\linewidth]{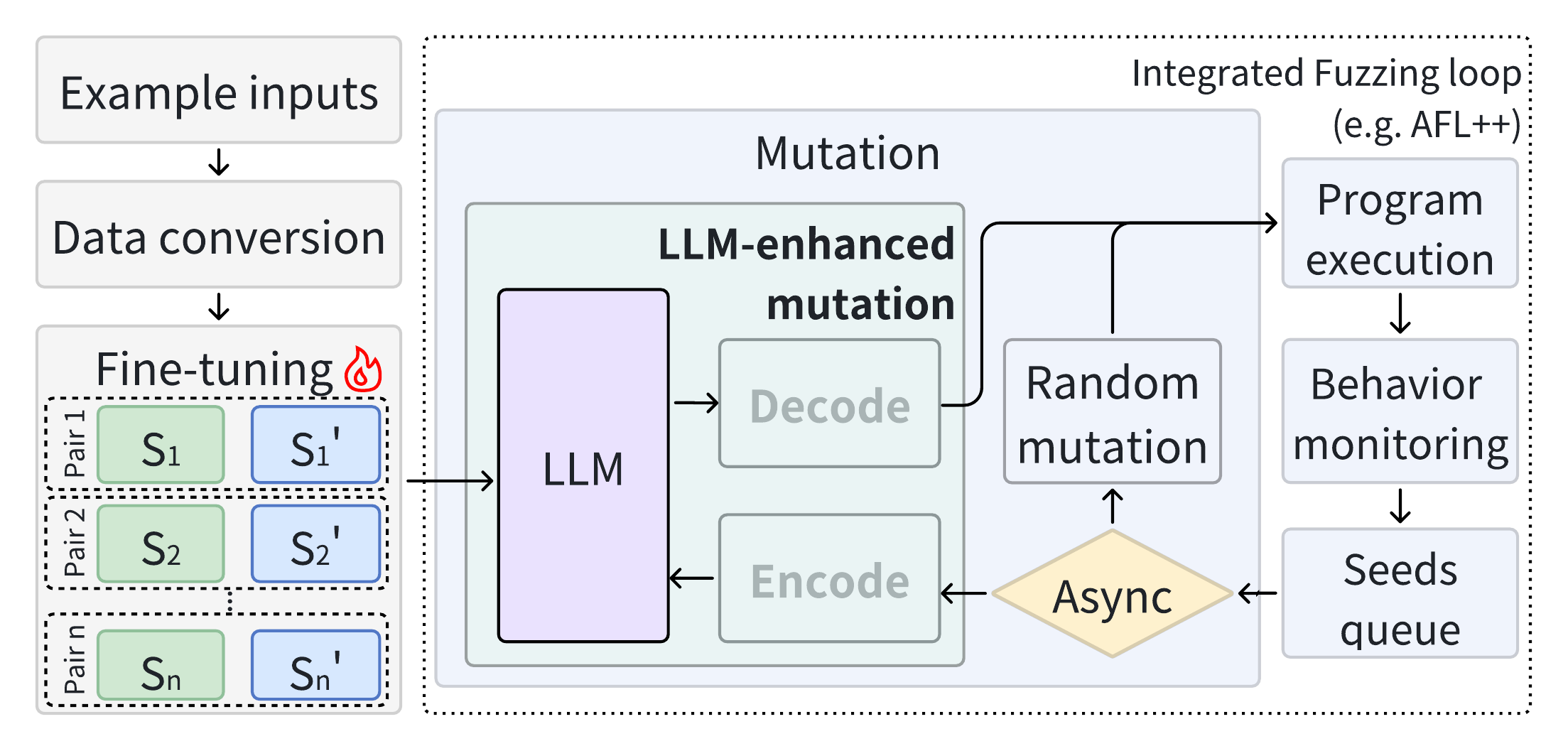}
      \caption{The overview of fuzzing with {LLAMAFUZZ}. 
      Each fine-tuning sample is a pair ($S_i$, ${S_i}'$), representing a seed before and after successful mutation. {LLAMAFUZZ} applies dual-layered, asynchronous mutations—traditional fuzzing and LLM-enhanced. The four light-blue boxes on the right (Execution, Behavior Monitoring, Mutation, Seeds Queue) depict the fuzzing loop.
      }
      \Description{The overview of fuzzing with {LLAMAFUZZ}. 
      Each fine-tuning sample is a pair ($S_i$, ${S_i}'$), representing a seed before and after successful mutation. {LLAMAFUZZ} applies dual-layered, asynchronous mutations—traditional fuzzing and LLM-enhanced. The four light-blue boxes on the right (Execution, Behavior Monitoring, Mutation, Seeds Queue) depict the fuzzing loop.}
      \label{architecture}
\vspace{-0.1cm}
\end{figure}

Fuzzer developers face a trade-off between employing general-purpose fuzzers and specialized ones. General-purpose fuzzers, while versatile, often struggle with handling structured seeds effectively. Meanwhile, specialized fuzzers can produce high-quality structured seeds, but this specialization can limit their flexibility and applicability.
Moreover, relying on grammar rules for seed generation requires extensive domain knowledge, which can be a barrier to their widespread use.

To address the aforementioned challenges, we propose an LLM-enhanced mutation strategy applicable to both binary and text-based formats with minimal fine-tuning. \autoref{architecture} provides an overview of {LLAMAFUZZ} architecture. Pre-trained on diverse datasets, LLMs can learn intricate patterns for data conversion and data format, which are crucial for structured data mutation. We further fine-tune LLMs to learn specific seed patterns and mutate structured seeds, aiming for a balance between generic and specialized fuzzers. As illustrated in \autoref{tab:xml_variants}, random mutations often perform low-level bit flips that corrupt the syntactic structure of valid inputs. In contrast, {LLAMAFUZZ} produces structurally valid and semantically meaningful variants, enabling more effective and targeted fuzzing.

We integrate LLMs and fuzzer with a lightweight asynchronous job queue, allowing {LLAMAFUZZ} to run efficiently on single or multiple GPUs. 
To assess bug-finding capabilities, we compared {LLAMAFUZZ} against state-of-the-art fuzzers on bug-based benchmark Magma~\cite{Hazimeh:2020:Magma}, including AFL++, Moptafl, Honggfuzz, and Fairfuzz. {LLAMAFUZZ} outperformed its top competitors, discovering 41 bugs on average and 47 unique bugs overall.
In addition, we evaluated {LLAMAFUZZ} on real-world programs across various structured data formats to access its versatility~\cite{FuzzBench}. {LLAMAFUZZ} performs competitively with grammar-based fuzzers and outperforms AFL++ on 11 out of 15 fuzzing targets.

\begin{table}[ht]
    \centering
    \scriptsize
    \renewcommand{\arraystretch}{1.12}
    \setlength{\tabcolsep}{2pt}

    \caption{
    Mutation example on XML.
    Format corruptions are highlighted in
    {\sethlcolor{lightred}\hl{red}},
    while successful and structurally valid mutations are highlighted in
    {\sethlcolor{lightgreen}\hl{green}}.
    }

    \resizebox{\linewidth}{!}{
    \begin{tabular}{>{\raggedright\arraybackslash}p{0.26\linewidth}
                    >{\raggedright\arraybackslash}p{0.26\linewidth}
                    >{\raggedright\arraybackslash}p{0.26\linewidth}}
        \hline
        \textbf{Sample Input} & \textbf{Random Mutation} & \textbf{{LLAMAFUZZ} Mutation} \\
        \hline

        \ttfamily
        \begin{tabular}[t]{@{}l@{}}
<doc> \\
\hspace*{0.5em} <message> \\
\hspace*{1.5em} <to>ace</to> \\
\hspace*{1.5em} <from>tom</from> \\
\hspace*{1.5em} <head>Reminder\\
\hspace*{2.5em} </head> \\
\hspace*{1.5em} <body>This is a \\
\hspace*{2.5em} reminder!</body>\\
\hspace*{0.5em} </message> \\
</doc>
        \end{tabular}
        &

        \ttfamily
        \begin{tabular}[t]{@{}l@{}}
<doc> \\
\hspace*{0.5em} <message> \\
\hspace*{1.5em} <to>ace<to> \\
\hspace*{1.5em} <from>tom{\sethlcolor{lightred}\hl{</from}} \\
\hspace*{1.5em} <head>Reminder\\
\hspace*{2.5em} </head> \\
\hspace*{1.5em} <body>A rem!nder!\\
\hspace*{2.5em} {\sethlcolor{lightred}\hl{</bod>}} \\
\hspace*{0.5em} {\sethlcolor{lightred}\hl{</messag>}} \\
</doc>
        \end{tabular}
        &

        \ttfamily
        \begin{tabular}[t]{@{}l@{}}
{\sethlcolor{lightgreen}\hl{<?xml version="1.0"}}\\
\hspace*{1.0em} {\sethlcolor{lightgreen}\hl{encoding="UTF-8"?>}} \\
<doc> \\
\hspace*{0.5em} <message> \\
\hspace*{1.5em} <to>{\sethlcolor{lightgreen}\hl{alex}}</to> \\
\hspace*{1.5em} <from>{\sethlcolor{lightgreen}\hl{ben}}</from> \\
\hspace*{1.5em} <head>{\sethlcolor{lightgreen}\hl{Hello}}</head> \\
\hspace*{1.5em} <body>This is a \\
\hspace*{3.5em} {\sethlcolor{lightgreen}\hl{hello}}!</body> \\
\hspace*{0.5em} </message> \\
\hspace*{0.5em} {\sethlcolor{lightgreen}\hl{<message>}} \\
\hspace*{1.5em} {\sethlcolor{lightgreen}\hl{<to>ben</to>}} \\
\hspace*{1.5em} {\sethlcolor{lightgreen}\hl{<from>alex</from>}} \\
\hspace*{1.5em} {\sethlcolor{lightgreen}\hl{<head>Reply:}}\\
\hspace*{3.5em} {\sethlcolor{lightgreen}\hl{Hello</head>}} \\
\hspace*{0.5em} {\sethlcolor{lightgreen}\hl{</message>}} \\
</doc>
        \end{tabular} \\
        \hline
    \end{tabular}
    }

    \label{tab:xml_variants}
    \vspace{-0.1cm}
\end{table}

\section{Background}

Greybox coverage-guided fuzzing tools~\cite{afl,fioraldi2020afl++,xu2024graphuzz,pham2019smart,honggfuzz} employ mutation-based strategies to explore input spaces; however, their reliance on random bit-level mutations often leads to inefficiencies and difficulties in generating valid inputs. To mitigate this, these tools use bitmaps to record execution paths and guide mutations toward unexplored code~\cite{wang2019sensitive}, although traditional bit-level approaches still struggle with highly structured data.
Grammar-based fuzzing~\cite{srivastava2021gramatron,blazytko2019grimoire,fioraldi2020weizz,aschermann2019nautilus} overcomes these challenges by using human-specified grammars and operators to generate syntactically valid and diverse inputs.

Recent advances with LLMs offer a promising alternative by effectively capturing complex data structures to guide seed mutations~\cite{deng2023large, xia2024fuzz4all, xia2023universal, huang2024large, yang2024whitefox}, as demonstrated by CHATFUZZ~\cite{hu2023augmenting} and compressed-language models~\cite{perez2024compressed}, thereby enhancing both efficiency and bug discovery in fuzzing. Our method leverages LLMs to learn valid input structures, enabling dynamic, structure-preserving mutations. This approach boosts mutation efficiency, code coverage, and bug detection compared to grey-box and grammar-based techniques.

\section{Method}

\subsection{Architecture}

We introduce {LLAMAFUZZ}, an LLM-enhanced greybox fuzzer designed to mutate structured data efficiently. As illustrated in \autoref{architecture}, our approach consisted of two primary stages. First, paired structured data was pre-processed~(\autoref{dataconversion}) and used to fine-tune the LLM, enabling LLMs to learn the underlying structural patterns and mutation transformation. Second, we integrated the fuzzer with LLM, which generated structured seeds based on existing inputs.

Our workflow included three parts:
\begin{enumerate*}[label=\textbf{\arabic*.}]
\item \textbf{Fine-tune preparation:} We collected training data from diverse sources. Also, we introduced a data conversion method that enables the LLM to mutate various data formats.
\item \textbf{Fine-tuning LLM for mutation:} We fine-tuned the LLM to perform structure-aware mutations.
\item \textbf{Integrate fuzzer and LLM:} We integrated the fuzzer with the LLM via an asynchronous communication mechanism.
\end{enumerate*}

Notably, we excluded seeds used in evaluation from the fine-tuning datasets to preempt potential data leakage,
limiting the possibility that the LLM replays memorized seeds to trigger bugs.

\subsection{Fine-tune preparation}
We followed the LLMs training process by generative pre-training of a language model on a diverse corpus of unlabeled text, followed by discriminative fine-tuning on specific tasks~\cite{lecun2015deep}. The fine-tuning data were collected from real-world fuzzing processes~\cite{FuzzBench}. We used these data to teach LLM the pattern and mutation of structured data, enabling LLM to modify a given seed to generate valuable seeds while keeping the original structure.

\paragraph{\textbf{Fine-tuning data collection}}
We expected the LLM to be able to understand the structure of data and generate structured seeds for testing, thus we needed to collect a training set first. Specifically, we collected valuable seeds from FuzzBench~\cite{FuzzBench} history experiment data and AFL++ fuzzing data that
\begin{enumerate*}[label=(\arabic*)]
    \item found new paths,
    \item had different hit-counts,
    \item triggered crashes.
\end{enumerate*}
The reasons are intuitive: improving coverage will help fuzzer explore target programs to find vulnerabilities in the unvisited path since bugs can not be found in undiscovered paths. While seeds with different hit counts\footnote{number of times a seed exercises a specific path} may not directly improve coverage, they execute the program in varied ways, potentially uncovering vulnerabilities in already visited paths. Ultimately, the goal of fuzzing is to find vulnerabilities, so any seed that triggers crashes is valuable.

\begin{figure}[ht]
     \centering
       \includegraphics[width=\linewidth]{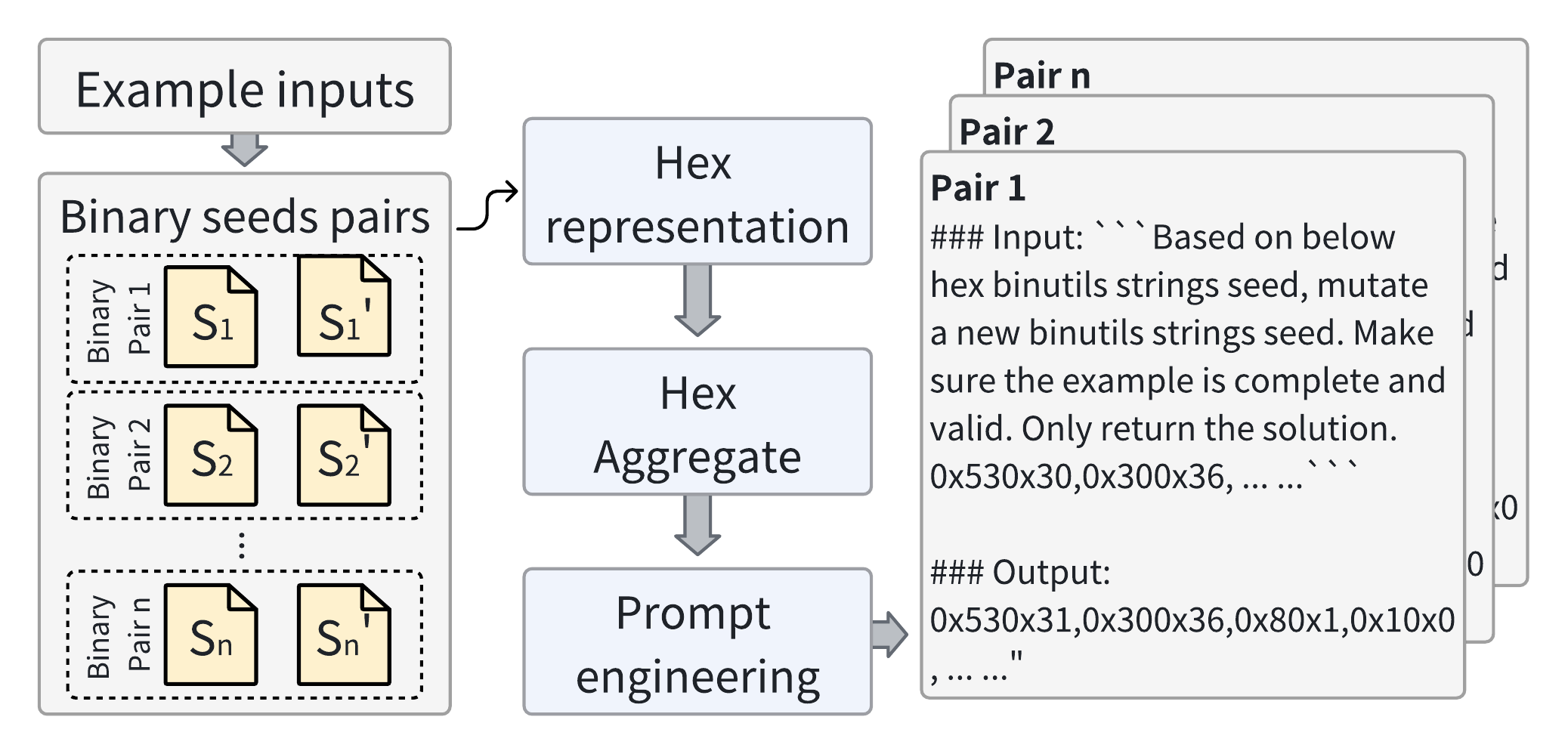}
      \caption{The workflow of dataset pre-processing. Each binary seed pair consists of the original seed $S_1$ and its mutated version ${S_{1}}'$. Pairs $1$ to $n$ on the right represent entries in the fine-tuning dataset.}
      \Description{The workflow of dataset pre-processing. Each binary seed pair consists of the original seed $S_1$ and its mutated version ${S_{1}}'$. Pairs $1$ to $n$ on the right represent entries in the fine-tuning dataset.}
      \label{dataconversion}
\vspace{-0.4cm} 
\end{figure}

\paragraph{\textbf{Data conversion and pre-processing}}
We construct a generic seed mutation model by converting binary input files to a uniform hexadecimal representation following \cite{perez2024compressed}. This conversion serves three purposes. First, to enable {LLAMAFUZZ} to handle various data formats, we need a uniform data reading method, as adapting to each format is impractical. Second, traditional fuzzers operate at the bit level on binary seeds, whereas LLMs typically require natural language input. Therefore, it is essential to convert the training data to a format that LLMs can understand. Third, the conversion is expected to be efficient, as delay would directly impact fuzzing throughput. Compared to other encoding schemes like base64, hexadecimal is more intuitive, easier to convert from binary. Note that this conversion applies only to binary data; for text-based data, we only add prompts to the seeds, as CHATFUZZ~\cite{hu2023augmenting} has shown LLM can process them directly.\label{Dataconversion}

As shown in \autoref{dataconversion}, our data conversion process involves three steps. Initially, the binary seed file is converted into a hexadecimal representation.  Next, every two contiguous hexadecimal characters are combined into a single unit. This approach not only reduced the token length of the input string, which was crucial given the limited maximum input token length of most current LLMs but also minimized the need to add new vocabulary to the tokenizer, compared to combining three or more hex characters. Finally, we add a prompt to each fine-tuning instance. In addition to data conversion, we incorporate noise data into our training set to mitigate overfitting and training data replay. Each training example consists of a pair of seeds: the original seed and its corresponding mutated version. This setup enables the LLM to learn both the mutation transformation and the underlying structure of the data formats.
    
\paragraph{\textbf{Tokenization}}
As aforementioned, each pair of contiguous hexadecimal characters was compiled into a phrase. However, not all such phrases were present in the tokenizer’s vocabulary. To tackle this, we employed a tokenizer that uses the Byte-Pair Encoding algorithm~\cite{sennrich2015neural,touvron2023llama}, which splits unknown phrases into smaller subword units. This allows the {LLAMAFUZZ} to generate and understand various data formats.

\subsection{Fine-tuning LLM for mutation}
    Fine-tuning pre-trained models is a common paradigm for achieving proficiency in specific downstream tasks~\cite{lecun2015deep,vonwerra2022trl,ding2023parameter,han2024parameter}. This process builds upon the pre-trained model's general understanding and adapts it to specific tasks through supervised learning. Similarly, supervised fine-tuning~(SFT) is necessary when employing a general-purpose LLM for structured data mutation.

    As shown in \autoref{dataconversion}, Pair $1$ to Pair $n$ provides example prompts to guide the model in structured data mutation. Each prompt consists of the original structured data and its desired mutated result in hexadecimal representation. Subsequently, we prompt the format keywords allowing LLM to take the general understanding of the format from pre-trained knowledge to do the mutation. 
    
    Following prior works~\cite{xia2024fuzz4all,yang2024fuzzcoder}, we apply SFT~\cite{vonwerra2022trl} to adapt the LLM. The supervised fine-tuning objective is defined as:
    \begin{equation}
    \mathcal{L}(\theta) = -\frac{1}{N}\sum_{i=1}^{N} \log P\Big( y^{(i)} \mid x^{(i)}; \theta \Big)
    \end{equation}
    where \(x^{(i)}\) denotes the i-th original seed, \(y^{(i)}\) its corresponding target mutated seed, and \(\theta\) represents the model parameters. The model is trained to generate \(y^{(i)}\) conditioned on \(x^{(i)}\).
    
    Additionally, we apply model quantization \cite{polino2018model} with mixed-precision $fp16$ and integrate LoRA~\cite{hu2021lora} to speedup training and inference while maintaining accuracy. 

\subsection{Integrate Fuzzer and LLM}
\label{Integrate fuzzer and LLM}
Speed is paramount for greybox fuzzers, which can execute hundreds or even thousands of seeds per second~\cite{yun2018qsym,zheng2019firm}. Any additional processes integrated into the greybox fuzzer, could potentially impair overall throughput and negatively impact fuzzing performance. Particularly, LLM generation is slower and more resource-intensive, primarily requiring substantial GPU resources.

To address this speed mismatch, we designed an asynchronous job queue for communication between the LLM and the fuzzer. The fuzzing process is formulated as follows:
\begin{equation}
    \begin{aligned}
    Q_{\mathrm{LLM}} &= \{(x^{(i)}, t_i) \mid x^{(i)} \in X,\, t_i \in T\}, \\
    Q_{\mathrm{AFL}} &= \{(x^{(i)}, e_i) \mid x^{(i)} \in X,\, e_i \in E\}, \\
    F(t) &= \begin{cases}
    F_{\mathrm{LLM}}(x^{(i)}) & \text{if } Q_{\mathrm{LLM}} \neq \emptyset, \\
    F_{\mathrm{AFL}}(x^{(i)}) & \text{otherwise}
    \end{cases}
    \end{aligned}
\end{equation}
where \(x^{(i)} \in X\) denotes the current seed test input chosen by the fuzzer, sets $T$ and $E$ represent the running times of LLM-generated and AFL-generated mutations. \(Q_{\mathrm{LLM}}\) and \(Q_{\mathrm{AFL}}\) are the asynchronous job queues for LLM mutation and AFL++ fuzzing respectively, with \(t_i\) and \(e_i\) representing their timestamps. The function \(F(t)\) defines the dual-layer fuzzing strategy. The seed selection process prioritizes test cases from both queues according to their coverage impact\footnote{The coverage function is an abstraction of behavior monitoring in the fuzzing loop, determining whether the current seed: (1) triggers new execution paths, (2) yields different hit-counts, or (3) results in crashes.}:
\begin{equation}
x^{i+1} = \max_{x \in  Q_{\text{LLM}} \cup Q_{\text{AFL}}}\, \{ \text{coverage}(x)\}
\end{equation}

This asynchronous process eliminates waiting time, enabling the fuzzer to run at high speed without delays from LLM mutation tasks. This ensures that integrating the LLM enhances the fuzzer's capabilities without compromising efficiency. Additionally, the dual-layer structure allows easy replacement of different LLMs.

\subsection{Asynchronous Queue Design}
        
        As summarized in Algorithm \ref{alg:hybrid_fuzzer}, our hybrid fuzzer orchestrates three concurrent components: Main Fuzzing Loop, LLM Mutation Thread, and AFL Mutation Thread.  This design decouples resource‐intensive, semantics‐driven mutations from the high‐throughput fuzzing loop, enabling non‐blocking operation and easy scaling or replacement of the LLM backend.

        \paragraph{\textbf{Main Fuzzing Loop.}}
        The Main Fuzzing Loop continuously selects and executes mutated seeds.  Whenever the LLM queue (Queue\_LLM) contains entries, those seeds take precedence; otherwise, it falls back to the AFL queue (Queue\_AFL).  Each seed is used to drive the target program, and a behavior monitor evaluates the execution trace.  Seeds that trigger new or anomalous behavior are deemed \textit{interesting} and are re‐enqueued into both Queue\_LLM\_unmutated and Queue\_AFL\_unmutated, ensuring they undergo further rounds of LLM‐ and AFL‐based mutation.
        
        \paragraph{\textbf{LLM Mutation Thread.}}
        This thread operates asynchronously on GPU resources. It dequeues seeds marked for LLM-based mutation, converts them into a hexadecimal representation, and utilizes a fine-tuned LLM to generate structured mutations. The resulting mutated seeds are then added back to the main LLM queue. This separation allows for computationally intensive LLM mutations to occur without blocking the faster main fuzzing loop.
        
        \paragraph{\textbf{AFL Mutation Thread.}}
        Running asynchronously on CPU resources, this thread handles AFL-based mutations. It dequeues seeds designated for AFL mutation, applies AFL's mutation strategies, and enqueues the resulting seeds into the AFL queue.
        
        By isolating CPU‐bound AFL work from GPU‐bound LLM tasks, the framework preserves the high throughput of the fuzzing pipeline while benefiting from both mutation paradigms.

    \begin{algorithm}[t]
    \caption{Hybrid Fuzzer with LLM- and AFL-based Mutations}
    \label{alg:hybrid_fuzzer}
    
    \KwIn{Initial seed corpus}
    \KwOut{Discovered interesting test cases}
    
    \BlankLine
    \textbf{Initialization:} start\_fuzzer(), llm\_mutate\_thread(), afl\_mutate\_thread()
    
    \BlankLine
    \SetKwFunction{Fuzzer}{start\_fuzzer}
    \SetKwFunction{LLMThread}{llm\_mutate\_thread}
    \SetKwFunction{AFLThread}{afl\_mutate\_thread}
    
    \BlankLine
    \Fuzzer{}{
        \While{true}{
            \eIf{Queue\_LLM not empty}{
                seed $\leftarrow$ dequeue(Queue\_LLM)
            }{
                seed $\leftarrow$ dequeue(Queue\_AFL)
            }
            \If{run\_fuzzer(seed) is interesting}{
                enqueue(Queue\_LLM\_unmutated, seed)\;
                enqueue(Queue\_AFL\_unmutated, seed)\;
            }
        }
    }
    
    \BlankLine
    \LLMThread{}{
        \While{true}{
            \If{Queue\_LLM\_unmutated not empty}{
                seed $\leftarrow$ dequeue(Queue\_LLM\_unmutated)\;
                hex\_seed $\leftarrow$ convert\_to\_hex(seed)\;
                mutated\_seed $\leftarrow$ LLM\_generate(hex\_seed)\tcp*[r]{GPU}
                enqueue(Queue\_LLM, mutated\_seed)\;
            }
        }
    }
    
    \BlankLine
    \AFLThread{}{
        \While{true}{
            \If{Queue\_AFL\_unmutated not empty}{
                seed $\leftarrow$ dequeue(Queue\_AFL\_unmutated)\;
                mutated\_seed $\leftarrow$ AFL\_generate(seed)\tcp*[r]{CPU}
                enqueue(Queue\_AFL, mutated\_seed)\;
            }
        }
    }
    \end{algorithm}

\section{Experiment}

%


\subsection{Benchmarks and Evaluation Metrics}
\paragraph{\textbf{Magma V1.2.}}\cite{Hazimeh:2020:Magma} is a ground-truth fuzzing benchmark suite based on real programs with real bugs.   \autoref{tab:benchmark} outlines details of fuzz targets, where four columns indicate benchmark, project, fuzz target, and expected file format. 
    In the Magma experiment, we compared {LLAMAFUZZ} with AFL++, Moptafl, Honggfuzz, and Fairfuzz.
    All baseline fuzzers except for AFL++\footnote{We used a more recent version of AFL++~(version 61e27c6) than the one provided in Magma, ensuring that we have access to the latest enhancements.} were provided in Magma. Following Magma setup, we use the number of detected bugs and the time to trigger them as key metrics.
    
    

    \paragraph{\textbf{OSS-Fuzz.}}
    We evaluated a set of real-world programs from OSS-Fuzz~\cite{serebryany2017oss}. For fairness and consistency in the evaluation process~\cite{klees2018evaluating}, we evaluated in a standard benchmark, FuzzBench~\cite{FuzzBench}. 
    The specific applications are detailed in \autoref{tab:benchmark}. The chosen benchmark encompassed 12 open-source programs that process different structured data in their latest versions. As suggested by \citet{klees2018evaluating} and \citet{Hazimeh:2020:Magma}, each experiment were repeated ten times, each trial lasted for 24 hours.
    We evaluated code coverage using established measures~\cite{bohme2022reliability,klees2018evaluating,wei2022free}, ensuring consistency by reporting branch coverage via afl-cov. Statistical significance was analyzed using the Mann-Whitney U test~\cite{klees2018evaluating}, and the Vargha-Delaney statistic~\cite{olsthoorn2020generating} quantified effect size.

    \paragraph{\textbf{Experiment Setup and Metrics.}}
 \begin{table}[htbp!]
    \centering
    \caption{Targets information. The programs tested under Magma are utilized in their default versions. For programs included in the real-world benchmark, we specify the exact versions used by listing the Git SHA identifiers.}
    \small
    \setlength{\tabcolsep}{4pt}
    \renewcommand{\arraystretch}{1.15}

    \begin{tabular}{%
        p{0.015\textwidth}  
        p{0.11\textwidth}  
        p{0.16\textwidth}  
        p{0.115\textwidth}  
    }
        \toprule
        & \textbf{Project \& version} & \textbf{Fuzz Target} & \textbf{File format} \\
        \midrule

        \multirow{21}{*}{\rotatebox[origin=c]{90}{\textbf{Magma V1.2}}}
        & libpng & libpng\_read\_fuzzer & PNG \\
        \cmidrule(lr){2-4}

        & libsndfile & sndfile\_fuzzer & Audio \\
        \cmidrule(lr){2-4}

        & \multirow{2}{*}{libtiff} & tiff\_read\_rgba\_fuzzer & \multirow{2}{*}{TIFF} \\
        &  & tiffcp &  \\
        \cmidrule(lr){2-4}

        & \multirow{2}{*}{libxml2} & read\_memory\_fuzzer & \multirow{2}{*}{XML} \\
        &  & xmllint &  \\
        \cmidrule(lr){2-4}

        & lua targets & lua & Lua \\
        \cmidrule(lr){2-4}

        & \multirow{2}{*}{openssl} & asn1, asn1parse, bignum, & \multirow{2}{*}{Binary blobs} \\
        &  & server, client, x509 &  \\
        \cmidrule(lr){2-4}

        & \multirow{4}{*}{php} & json & JSON \\
        &  & exif & EXIF \\
        &  & unserialize & Serialized object \\
        &  & parser & PHP \\
        \cmidrule(lr){2-4}

        & \multirow{2}{*}{poppler} & pdf\_fuzzer, pdfimages & \multirow{2}{*}{PDF} \\
        &  & pdftoppm &  \\
        \cmidrule(lr){2-4}

        & sqlite3 & sqlite3\_fuzz & SQL query \\
        \cmidrule(lr){1-4}

        \multirow{20}{*}{\rotatebox[origin=c]{90}{\textbf{Real-world programs}}}
        & mruby 14c2179 & mruby\_fuzzer & mruby \\
        \cmidrule(lr){2-4}

        & php afa034d & php\_exec & PHP \\
        \cmidrule(lr){2-4}

        & quickjs 91459fb & eval & JavaScript \\
        \cmidrule(lr){2-4}

        & \multirow{2}{*}{binutils 7320840} & fuzz\_nm, fuzz\_objcopy & \multirow{2}{*}{ELF, String} \\
        &  & fuzz\_readelf, fuzz\_strings &  \\
        \cmidrule(lr){2-4}

        & bloaty 34f4a66 & fuzz\_target & ELF, Mach-O \\
        \cmidrule(lr){2-4}

        & freetype2 cd02d35 & ftfuzzer & TTF, OTF, WOFF \\
        \cmidrule(lr){2-4}

        & grok b9286c2 & decompress\_fuzzer & JPEG 2000 \\
        \cmidrule(lr){2-4}

        & \multirow{2}{*}{kamailio 3f774f3} & fuzz\_parse\_msg & SIP message \\
        &  & fuzz\_uri & URI \\
        \cmidrule(lr){2-4}

        & \multirow{3}{*}{libavc 828cdb7} & avc\_dec\_fuzzer & AVC \\
        &  & mvc\_dec\_fuzzer & MVC \\
        &  & svc\_dec\_fuzzer & SVC \\
        \cmidrule(lr){2-4}

        & libhevc d0897de & hevc\_dec\_fuzzer & HEVC \\
        \cmidrule(lr){2-4}

        & openh264 1c23887 & decoder\_fuzzer & H.264 / MPEG-4 AVC \\
        \cmidrule(lr){2-4}

        & zlib 0f51fb4 & uncompress\_fuzzer & Zlib compressed \\
        \bottomrule
    \end{tabular}

    \vspace{-0.2cm}
    \label{tab:benchmark}
\end{table}

    We chose Magma for several reasons. First, Magma involves a wide range of popular programs that process diverse structured data with real-world environments, including 9 libraries and 21 objects. Second, unlike LAVA-M~\cite{dolan2016lava}, which primarily employs synthetic bugs and magic byte comparisons, Magma offers a diverse range of real vulnerabilities, with a total of 138 bugs spanning integer errors, divide-by-zero faults, memory overflows, use-after-free, double-free, and null-pointer dereference scenarios. It incorporates real-world bugs from older versions of software updated to their latest releases, ensuring the benchmark’s relevance and practical applicability. Third, Magma focuses on bug counts and time-to-bug metrics as more direct surrogates for performance~\cite{Hazimeh:2020:Magma}. In addition, we chose FuzzBench in later real-world experiments, because it employed Docker containers to standardize the testing environment for each fuzzer. This setup guaranteed fairness that all fuzzers operated under identical conditions, thus ensuring comparability of results.

    \autoref{tab:benchmark} summarizes the fuzzing targets used in our experiments, covering both Magma and real-world programs. The Magma benchmark includes 9 projects, comprising 21 fuzz targets across 12 different file formats. The real-world benchmark features 12 projects, with 18 fuzz targets spanning 19 file formats. Our selection followed three criteria. First, the benchmark had to have a diverse structure format. Second, the program needed to handle complex structured data. Third, the program had to be popular and important.

    To assess the statistical significance of our results, we applied the Mann-Whitney U Test~\cite{klees2018evaluating}. As per the Mann-Whitney U-test, a result is statistically significant if the p-value is less than 0.05. In addition, we used the Vargha-Delaney statistic~\cite{olsthoorn2020generating} to quantify effect size, providing insight into the magnitude of observed differences. The result was classified as negligible if $\hat{A}_{12}$ was less than 0.5, as medium if it was greater than 0.5 but not exceeding 0.8, and as large if it was greater than 0.8.

\subsection{Implementation details}
    We implemented {LLAMAFUZZ} by extending AFL++~\cite{fioraldi2020afl++} to evaluate the potential of LLMs in addressing the limitations of traditional fuzzing on structured inputs. The base model used is LLaMA 2-7B-Chat~\cite{touvron2023llama}, chosen for its balance between efficiency and reasoning capability. Since {LLAMAFUZZ} is built on top of AFL++, any performance deviation can be attributed to our LLM-driven mutation module. To mitigate seed memorization, all experimental seeds were excluded from the fine-tuning dataset.

    We followed the official build procedures of OSS-Fuzz~\cite{serebryany2017oss}, Fuzz-Bench~\cite{FuzzBench}, and Magma~\cite{Hazimeh:2020:Magma} to prepare fuzzing targets. For OSS-Fuzz benchmarks, we used the default initial seed corpus provided by the project. For Magma benchmarks, we similarly adopted the benchmark-defined default seeds to ensure experimental consistency.

    \paragraph{\textbf{Fine-tuning LLM}}
        For fine-tuning, we utilized a dataset of approximately 10,000 samples per target. The base model, LLaMA 2-7B, was fine-tuned using LoRA (Low-Rank Adaptation) with an adaptation rank of 8, lora\_alpha set to 16, and a dropout rate of 0.05. The training was conducted over 20 epochs with a batch size of 1 per device and a maximum sequence length of 1400 tokens on a single A100 GPU. Overall, the fine-tuning process took around 6 hours per fuzzing target. We employed a cosine learning rate scheduler with a learning rate of 2e-4, 30 warm-up steps, and a weight decay of 0.001. Optimization was performed using AdamW. Gradient checkpointing was enabled to reduce memory usage, and 4-bit quantization with bfloat16 compute precision was applied for efficiency. To ensure consistency across runs, branch coverage was computed using afl-cov, and the final model checkpoint was saved for deployment.

    \paragraph{\textbf{System Integration}}
    As described in \autoref{Integrate fuzzer and LLM}, we developed an asynchronous job-queue system for communication between the fuzzer and the LLM. The AFL++ source code (e.g., afl-fuzz-bitmap.c) was modified to integrate {LLAMAFUZZ}’s key functionalities, including the message queue, seed evaluation, and mutation invocation mechanisms. Following prior work~\cite{radford2019language,wang2022self}, we used a temperature of 1.0 to ensure balanced exploration and factual mutation behavior. We further adopted mixed-precision quantization~\cite{polino2018model} (FP16) and parameter freezing via LoRA~\cite{hu2021lora} to accelerate both fine-tuning and inference with minimal degradation in accuracy.
    
\begin{figure}[ht]
     \centering
       \includegraphics[width=0.76\linewidth]{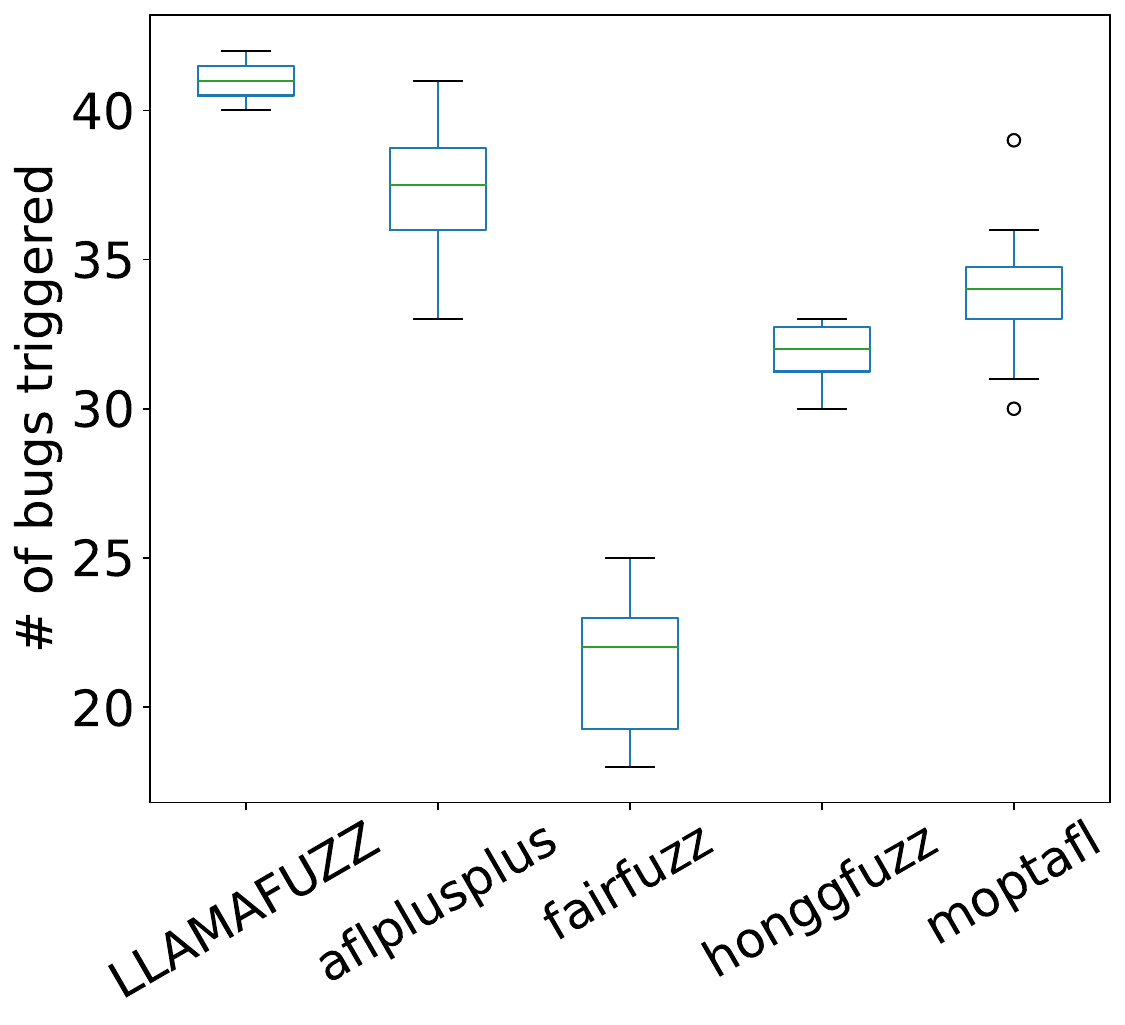}
     \caption{Distribution of the average number of bugs triggered over 24 hours.}
     \Description{Distribution of the average number of bugs triggered over 24 hours.}
     \label{box}
\vspace{-0.2cm}
\end{figure}

\begin{figure}[ht]
     \centering
       \includegraphics[width=\linewidth]{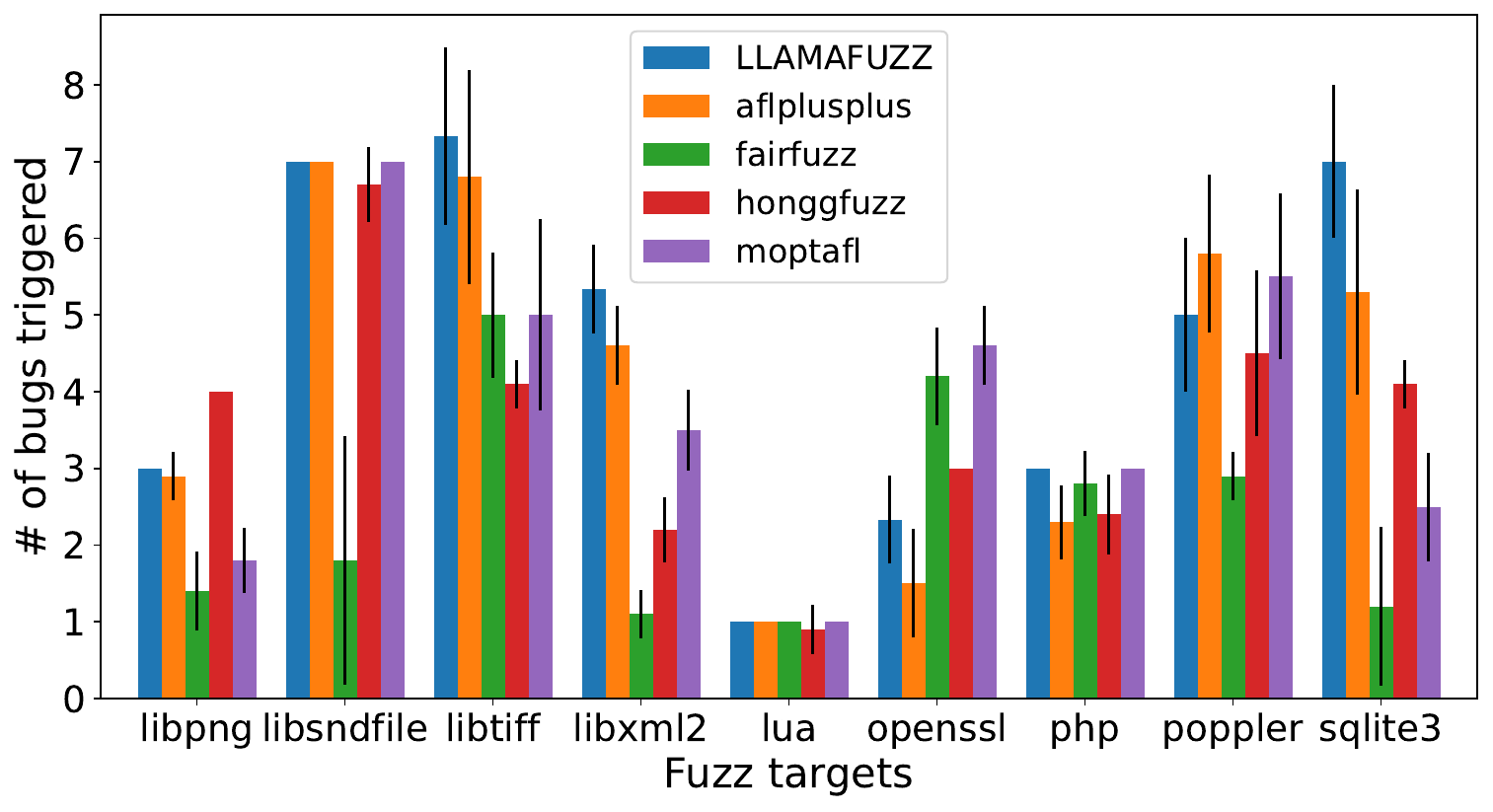}
    \caption{Arithmetic mean number of bugs identified for each project per trial. The black line denotes the 95\% confidence interval.}
    \Description{Arithmetic mean number of bugs identified for each project per trial. The black line denotes the 95\% confidence interval.}
    \label{eva}
\vspace{-0.2cm}
\end{figure}

\subsection{Main results}
\paragraph{\textbf{{LLAMAFUZZ} find more bugs}}
\label{exp1}

    

    We evaluated {LLAMAFUZZ} performance against other popular fuzzers by checking whether it is able to discover more bugs on Magma. 
    According to \autoref{box}, {LLAMAFUZZ} outperforms all other fuzzers in terms of the average number of bugs triggered for each trail. 
    As aforementioned, {LLAMAFUZZ} was built upon AFL++. Any improvement of {LLAMAFUZZ} over AFL++ can be attributed to the contribution of LLM.
    These results highlighted {LLAMAFUZZ}'s competitiveness and robustness with the SOTA in bug-triggering capabilities.

    To further investigate the performance of {LLAMAFUZZ} across different fuzzing targets, \autoref{eva} presents the arithmetic mean number of bugs identified for each project per trial per day. According to the results, {LLAMAFUZZ} triggered the most unique bugs among the evaluated fuzzers. It discovered 47 unique bugs in Magma, while AFL++, Moptafl, Honggfuzz, and Fairfuzz found 46, 42, 37, and 31 errors, respectively. Vulnerabilities were found in 9 tested implementations and encompassed various types of memory vulnerabilities, including use-after-free, buffer overflow, and memory leaks. Notably, SQL003, XML006, and XML002 were never found by any other fuzzers. 
    

    \begin{figure}[ht]
        \includegraphics[width=0.5\textwidth]{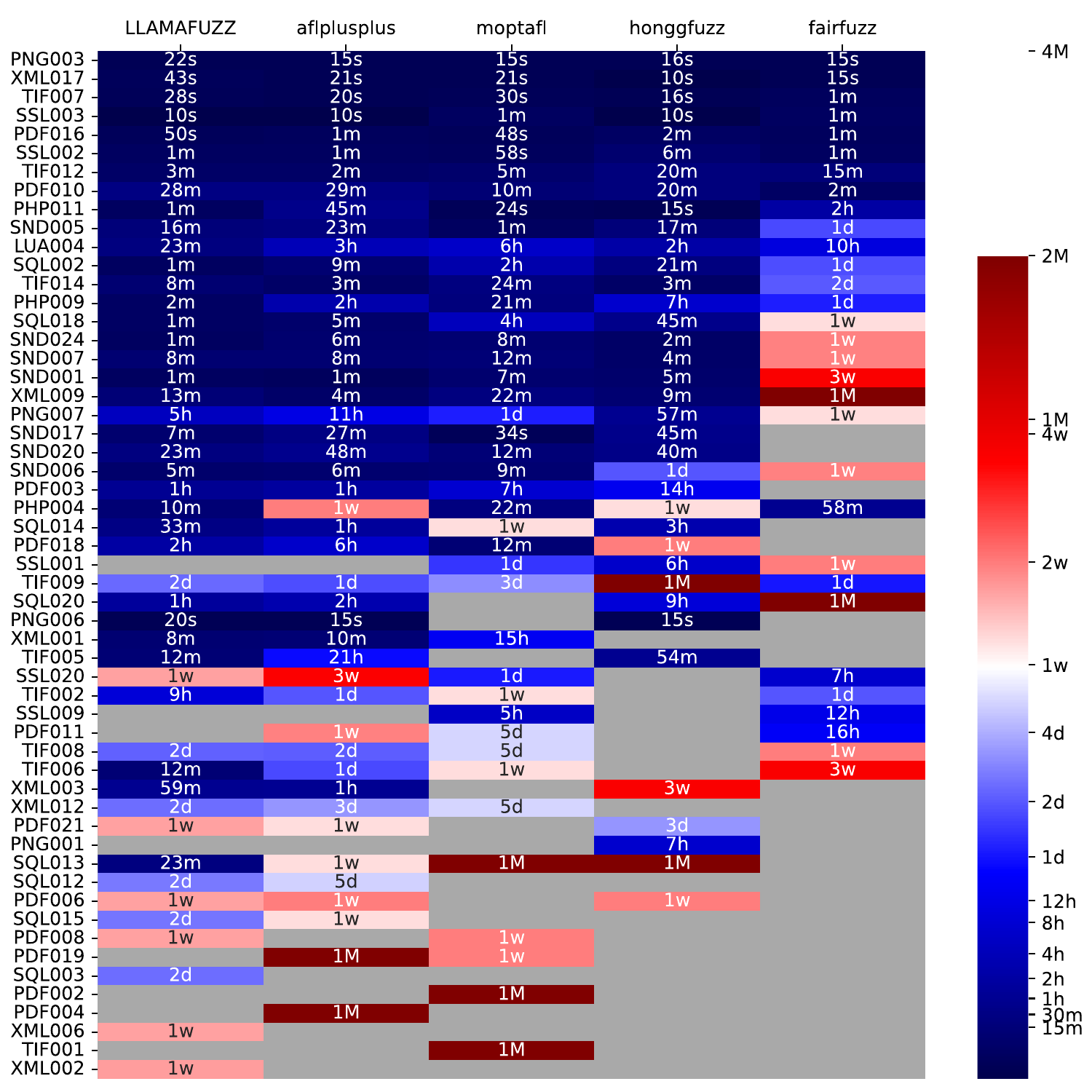}
        \caption{Heatmap of expected bug trigger time achieved by {LLAMAFUZZ}, AFL++, Moptafl, Honggfuzz, and Fairfuzz at the end of each 24-hour trail. Within each block, more intense blue shades denote shorter expected trigger times, while the grey parts represent bugs that were not triggered in any trials by that fuzzer.}
        \Description{Heatmap of expected bug trigger time achieved by {LLAMAFUZZ}, AFL++, Moptafl, Honggfuzz, and Fairfuzz at the end of each 24-hour trail. Within each block, more intense blue shades denote shorter expected trigger times, while the grey parts represent bugs that were not triggered in any trials by that fuzzer.}
        \label{bugs avg time}
        \vspace{-0.1cm}
    \end{figure}
    
    To understand the contributions of the LLM mutation, we conducted a more detailed investigation. Bug XML006 (CVE-2017-9048) demonstrated that randomness-only mutation was insufficient; a comprehensive understanding through structure is necessary. XML006 is a stack-based buffer overflow vulnerability in libxml2. To trigger it, the mutator must recursively dump the element content definition into a char buffer \textit{buf} of size \textit{size}. At the end of the routine, the mutator appended two more characters to exceed the \textit{size}. In our experiment, only {LLAMAFUZZ} triggered this bug, demonstrating that {LLAMAFUZZ} can facilitate the host fuzzers with the capability of finding bugs.
    


\paragraph{\textbf{Performance on bug-triggered time}}

    Consequently, we listed all the unique bugs triggered, including bug ID and the expected time taken to trigger it in \autoref{bugs avg time}. The reported time accounts for missed measurements (where the fuzzer only triggered a bug in M out of N campaigns) and fits the distribution of time-to-bug samples onto an exponential distribution~\cite{Hazimeh:2020:Magma}.

    Compared to AFL++, {LLAMAFUZZ} triggered a greater number of bugs and significantly sped up, with darker blue grid cells representing faster bug triggering. Specifically, {LLAMAFUZZ} achieved significant speedups in 29 of 43 bugs that were triggered in both {LLAMAFUZZ} and AFL++ with the remaining bugs exhibiting similar trigger times. In comparison to moptafl, honggfuzz, and fairfuzz, {LLAMAFUZZ} triggered bugs faster in 25, 23, and 21 cases respectively. Overall, the results indicate a substantial advantage of {LLAMAFUZZ} over AFL++, Moptafl, Honggfuzz, and Fairfuzz in exploring bugs.

\begin{table*}[htbp]
        \centering
        \fontsize{7.9}{7}\selectfont
        \caption{Average branch coverage achieved by our approach ({LLAMAFUZZ}) and the baseline (AFL++). We report the average branch coverage, p-values produced by the Mann-Whitney U Test, and the Vargha-Delaney statistics ($\hat{A}_{12}$). For assessing effect size, we use the labels -, M, and L to represent negligible, medium, and large effects, respectively.}
        \begin{tabular}{llccccc}
        \toprule
        \multirow{3}{*}{\textbf{Fuzz target}} & \multirow{3}{*}{\textbf{Fuzz object}} & \multicolumn{5}{c}{\textbf{Branch coverage~(avg)}}\\
        \cmidrule(lr){3-7}
         & & {LLAMAFUZZ} & AFL++ & Improv. & p-value & $\hat{A}_{12}$\\
        \midrule
        \multirow{4}{*}{binutils} & fuzz\_nm & \num{13969} & \num{9017} & \num{54.91}\% & <0.01 & L(1)\\
        & fuzz\_objcopy & \num{22118} & \num{12494} & 77.03\% & <0.01 & L(1)\\
        & fuzz\_readelf & \num{6552} & \num{4437} & 47.67\% & <0.01 & L(1)\\
        & fuzz\_strings & \num{6441} & \num{5295} & 21.64\% & <0.01 & L(1)\\
        \midrule
         bloaty & fuzz\_target & \num{5972} & \num{5722} & 4.37\% & <0.01 & L(1)\\
        \midrule
        freetype2 & freetype2-ftfuzzer & \num{10521} & \num{9978} & 5.45\% & <0.01 & L(1)\\
        \midrule
        grok & grk\_decompress\_fuzzer & \num{3721} & \num{2313} & 60.87\% & <0.01 & L(1)\\
        \midrule
        \multirow{2}{*}{kamailio} & fuzz\_parse\_msg & \num{3743} & \num{2692} & 39.06\% & 0.03 & L(0.95) \\
        & fuzz\_uri & \num{1392} & \num{1391} & 0.04\% & 0.72 & M(0.55)\\
        \midrule
        \multirow{3}{*}{libavc} & avc\_dec\_fuzzer & \num{9872} & \num{9838} & 0.35\% & 0.01 & L(1)\\
        & mvc\_dec\_fuzzer & \num{6463} & \num{5933} & 8.94\% & 0.87 & M(0.55)\\
        & svc\_dec\_fuzzer & \num{11212} & \num{6403} & 75.11\% & 0.08 & L(0.87)\\
        \midrule
        libhevc & hevc\_dec\_fuzzer & \num{15154} & \num{15122} & 0.21\% & 0.22 & L(0.77)\\
        \midrule
        openh264 & decoder\_fuzzer & \num{7394} & \num{7396} & -0.03\% & 0.79 & M(0.57)\\
        \midrule
         zlib& zlib\_uncompress\_fuzzer & 387 & 385 & 0.48\% &0.60 & -\\
        \bottomrule
        \end{tabular}
    \label{fuzzbench code coverage}
    \end{table*}
    
\subsection{Performance comparison on OSS-Fuzz}
\label{exp2}


    While performing well on Magma, we are committed to further validating its efficacy in real-world applications. To this end, we have selected a series of open-source programs to conduct a comprehensive evaluation. This step is crucial for demonstrating the practical effectiveness of {LLAMAFUZZ}'s methodologies and techniques in various file formats under real-world conditions. 

    First, we evaluated the performance of {LLAMAFUZZ} against specialized grammar-based fuzzers, including Gramatron~\cite{srivastava2021gramatron}, Grimoire~\cite{blazytko2019grimoire}, and Nautilus~\cite{aschermann2019nautilus}. We selected a common set of fuzzing targets compatible with all specialized grammar-based fuzzers, including mruby, PHP, and quickjs.
    Next, we conducted a more comprehensive evaluation using real-world programs. Given that grammar-based fuzzers are limited to well-defined, specialized targets, AFL++ was chosen as the reference competitor, representing the state-of-the-art in greybox fuzzing~\autoref{exp1}.

\begin{figure}[ht]
     \centering
       \includegraphics[width=\linewidth]{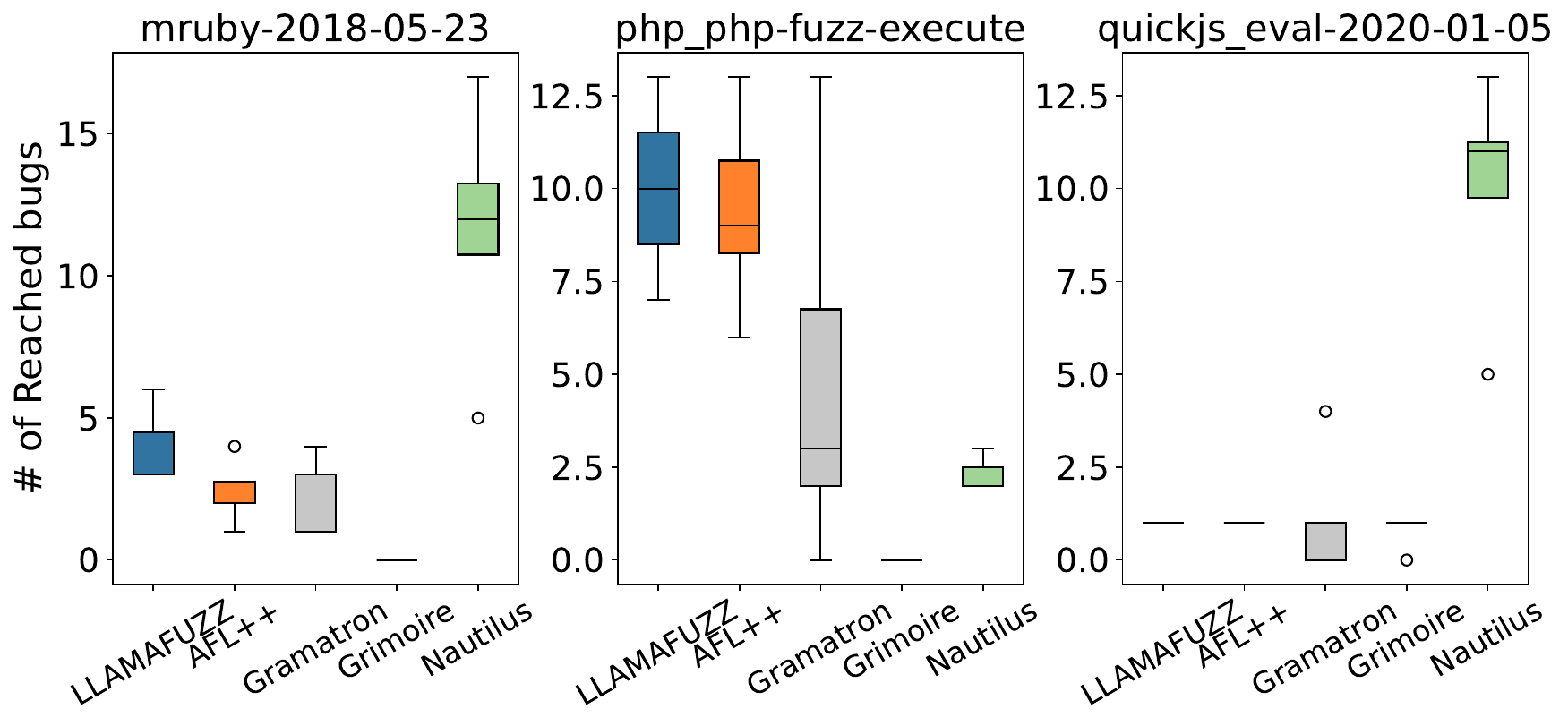}
         \caption{The distribution of reached bug achieved by {LLAMAFUZZ}, AFL++, Gramatron, Grimoire, and Nautilus.}
         \Description{The distribution of reached bug achieved by {LLAMAFUZZ}, AFL++, Gramatron, Grimoire, and Nautilus.}
         \label{grammar}
    \vspace{-0.2cm}
    \end{figure}

    \paragraph{\textbf{Compare with grammar-based fuzzers}}
    As illustrated in \autoref{grammar}, across all three targets, {LLAMAFUZZ} outperformed the baseline AFL++. Specifically, {LLAMAFUZZ} identified 13 bugs in PHP, ranking \#1 among all evaluated fuzzers. In the case of mruby, {LLAMAFUZZ} achieved \#2 rank. Notably, each selected grammar-based fuzzer demonstrated significant performance only on a subset of the fuzzing targets. For example, Nautilus excelled in mruby and quickjs but did not perform competitively on PHP. In contrast, {LLAMAFUZZ} showed a strong performance across all targets.

    \paragraph{\textbf{Compare with AFL++}}
    To conduct a more comprehensive evaluation of {LLAMAFUZZ} on real-world programs, we selected AFL++ as the reference competitor, representing the state-of-the-art in greybox fuzzing.

    \autoref{fuzzbench code coverage} reports the average branch coverage, the percentage improvement in average branch coverage over the same timeframe (see column \textbf{Improv}), p-value, and $\hat{A}_{12}$. In 11 out of 15 targets, {LLAMAFUZZ} shows statistically significant improvement compared with AFL++ in terms of code coverage.
    Furthermore, across all evaluated components, except for zlib, {LLAMAFUZZ} exhibited medium to large effect sizes. These results underscore {LLAMAFUZZ}'s effectiveness in enhancing coverage across a variety of applications.
    

    While performance on a few targets, such as zlib, kamailio-uri, and openh264, was comparable to AFL++, improvements were more modest or statistically less significant. These cases may reflect challenges in processing larger or more complex seeds, which exceed the LLM's context capacity in the fine-tuning data for domain-specific formats. We believe these results highlight opportunities to further enhance {LLAMAFUZZ} through better domain adaptation and LLM scaling.

\begin{table*}[htbp!]
        \centering
        \fontsize{7.9}{7}\selectfont
        \caption{Average branch coverage achieved by {LLAMAFUZZ}, baseline with default initial seeds (AFL++) and baseline with default initial seeds combined with trimmed seeds set from {LLAMAFUZZ}’s fine-tuning dataset (\textit{AFL++}${}^*$).}
        \begin{tabular}{llccccc}
        \toprule
        \multirow{5}{*}{\textbf{Fuzz target}} & \multirow{5}{*}{\textbf{Fuzz object}} & \multicolumn{4}{c}{\textbf{Branch coverage~(avg)}}\\
        \cmidrule(lr){3-7}
        & & \multirow{3}{*}{AFL++}& \multicolumn{4}{c}{ {LLAMAFUZZ} integrated }\\
        \cmidrule(lr){4-7}
         &  & & Llama2 & Llama3 & Mistral & Qwen2 \\
        \midrule
        binutils & fuzz\_nm & \num{9017} & \num{13958}  & \num{14423} & \num{14929} & \num{15041} \\
        \midrule
        grok & grk\_decompress\_fuzzer & \num{2313} & \num{3750} & \num{4028} & \num{3839} &\num{4123} \\
        \midrule
        kamailio & fuzz\_parse\_msg & \num{2692} & \num{3743} & \num{3821} & \num{3911} & \num{4013} \\
        \bottomrule
        \end{tabular}
    \label{tab:appenidx_llms}
\end{table*}

\begin{table}[h]
\centering
\scriptsize
\caption{Average branch coverage achieved by {LLAMAFUZZ} compared with without finetuning.}
\resizebox{\linewidth}{!}{%
  \begin{tabular}{llcccc}
        \toprule
        \multirow{4}{*}{\textbf{Fuzz target}} & \multirow{4}{*}{\textbf{Fuzz object}} & \multicolumn{4}{c}{\textbf{Branch coverage~(avg)}}\\
        \cmidrule(lr){3-6}
         &  & Llama2 & \multirow{2}{*}{Llama2} & Llama3 & \multirow{2}{*}{Llama3} \\
         &  &  (w/o finetune) & & (w/o finetune) & \\
        \midrule
        binutils & fuzz\_nm & \num{10247} & \num{13958}  & \num{12711} & \num{14423}  \\
        \midrule
        grok & grk\_decompress\_fuzzer & \num{3023} & \num{3750} & \num{3761} & \num{4028}  \\
        \midrule
        kamailio & fuzz\_parse\_msg & \num{3491} & \num{3743} & \num{3634} & \num{3821} \\
        \bottomrule
    \end{tabular}}
\label{tab:ablation fine-tune}
\vspace{-0.5cm}
\end{table}

\subsection{Ablation study: Generalization across different models}
\label{appendix:exp_diff_llms}
We selected a small set of real-world programs to demonstrate the generalization of our method across different LLMs. We select llama2-7b-chat-hf~\cite{touvron2023llama}, LLaMA-3-8B-Instruct\cite{dubey2024llama}, Mistral-7B-Instruct-v0.2\cite{jiang2023mistral}, and Qwen2-7B-Instruct\cite{yang2024qwen2} in this experiment to balance GPU cost. \autoref{tab:appenidx_llms} outlines the results, comparing the performance of our LLM-enhanced fuzzing approaches against the baseline AFL++. Across all evaluated targets, LLM-enhanced methods outperformed AFL++, achieving higher branch coverage. Notably, more recent LLMs, such as LLaMA3-8B, Mistral, and Qwen-7B, exhibited superior performance, likely due to their improved inference capabilities and better contextual understanding. The result demonstrates the generalizability of our method, confirming its adaptability to various LLMs while consistently enhancing fuzzing effectiveness.

\subsection{Ablation study: Effectiveness of fine-tuning}
To examine whether fine-tuning enhances the fuzzing capability of {LLAMAFUZZ}, we compare average branch coverage before and after mdoel fine-tuning. As illustrate in \autoref{tab:ablation fine-tune}, fine-tuning consistently improves coverage across all models and targets, confirming its effectiveness in adapting LLMs to fuzzing-specific objectives. 
These gains demonstrate that model adaptation enables {LLAMAFUZZ} to generate inputs more aligned with the syntactic and semantic structure of the fuzzing domain, leading to deeper code exploration. Notably, even the stronger Llama3 backbone benefits from task-specific fine-tuning, indicating that general-purpose instruction-tuning alone is insufficient for coverage-oriented fuzzing.

\section{Analysis}
    We have demonstrated the superiority of {LLAMAFUZZ} among standard benchmark and real-world programs. Moving forward, we first establish that the performance improvements of {LLAMAFUZZ} are attributable to LLM’s inference capabilities rather than the replay of fine-tuning data. Lastly, we explore how LLM enhances the fuzzing.

    \subsection{Did the LLM improve fuzzing by memorizing the fine-tuning data?}

    \begin{table*}[htbp!]
        \centering
        \fontsize{7.9}{7}\selectfont
        \caption{Average branch coverage achieved by {LLAMAFUZZ}, baseline with default initial seeds (AFL++) and baseline with default initial seeds combined with trimmed seeds set from {LLAMAFUZZ}’s fine-tuning dataset (\textit{AFL++}${}^*$).}
        \begin{tabular}{llcccccc}
        \toprule
        \multirow{3}{*}{\textbf{Fuzz target}} & \multirow{3}{*}{\textbf{Fuzz object}} & \multicolumn{5}{c}{\textbf{Branch coverage~(avg)}}\\
        \cmidrule(lr){3-8}
         & & {LLAMAFUZZ} & AFL++ & \textit{AFL++}${}^*$ & Improv. & p-value & $\hat{A}_{12}$\\
        \midrule
        \multirow{4}{*}{binutils} & fuzz\_nm & \num{13969} & \num{9017} & \num{10983} & 27.18\% & <0.01 & L(1)\\
        & fuzz\_objcopy & \num{22118} & \num{12494} & \num{13575} & 62.93\% & <0.01 & L(1)\\
        & fuzz\_readelf & \num{6552} & \num{4437} & \num{5789} & 13.79\% & <0.01 & L(1)\\
        & fuzz\_strings & \num{6441} & \num{5295} & \num{5220} & 23.39\% & <0.01 & L(1)\\
        \midrule
         bloaty & fuzz\_target & \num{5972} & \num{5722} & \num{5918} & 0.90\% & 0.37 & M(0.7)\\
        \midrule
        freetype2 & freetype2-ftfuzzer & \num{10521} & \num{9978} & \num{10838} & -2.92\% & 0.01 & -\\
        \midrule
        grok & grk\_decompress\_fuzzer & \num{3721} & \num{2313} & \num{2629} & 41.54\% & 0.02 & L(1)\\
        \midrule
        kamailio & fuzz\_parse\_msg & \num{3743} & \num{2692} & \num{2521} & 48.49\% & <0.01 & L(1) \\
        \midrule
        libavc & avc\_dec\_fuzzer & \num{9872} & \num{9838} & \num{9847} & 0.26\% & 0.01 & L(1)\\
        \bottomrule
        \end{tabular}
    \label{ablation code coverage}
    \end{table*}
    
    Although \autoref{exp1} and \autoref{exp2} demonstrate the effectiveness of {LLAMAFUZZ}, we need to investigate whether the performance gains are due to fine-tuning data replay. To address this, we conducted an ablation study focusing on the significant performance improvements observed in \autoref{exp2}. We used default initial seeds from OSS-FUZZ used by {LLAMAFUZZ} and compared it to the default initial seeds combined with trimmed seeds set from {LLAMAFUZZ}’s fine-tuning dataset as initial seeds for AFL++ (see column \textit{AFL++}${}^*$ in \autoref{ablation code coverage}). If {LLAMAFUZZ} were naively replaying fine-tuning data, we would expect \textit{AFL++}${}^*$ to outperform {LLAMAFUZZ} across all fuzzing targets.

    As detailed in \autoref{ablation code coverage}, {LLAMAFUZZ} significantly outperformed \textit{AFL++}${}^*$ across all targets except for bloaty and freetype2, demonstrating that LLM enhances the fuzzing process by interpreting structured data rather than simply replaying fine-tuning data. In the cases of freetype2 and bloaty, however, the performance was similar. 
    This is attributed to the fact that trimmed initial seeds set allowed \textit{AFL++}${}^*$ to achieve high coverage in the early fuzzing process, which introduces a bias in favor of \textit{AFL++}${}^*$. 
    Despite this, the comparable final performance demonstrates that LLM augments the fuzzing process through its intrinsic understanding and inference capabilities.

    Interestingly, we observed a slight performance decline in kamailio and binutils\_fuzz\_strings when comparing \textit{AFL++}${}^*$ with AFL++. This may due to the inherent randomness in the AFL++. On the other hand, this also indicates that the fine-tuning dataset collection has successfully avoided overlap with the testing data.


\begin{figure}[ht]
\centering
\includegraphics[width=0.84\linewidth]{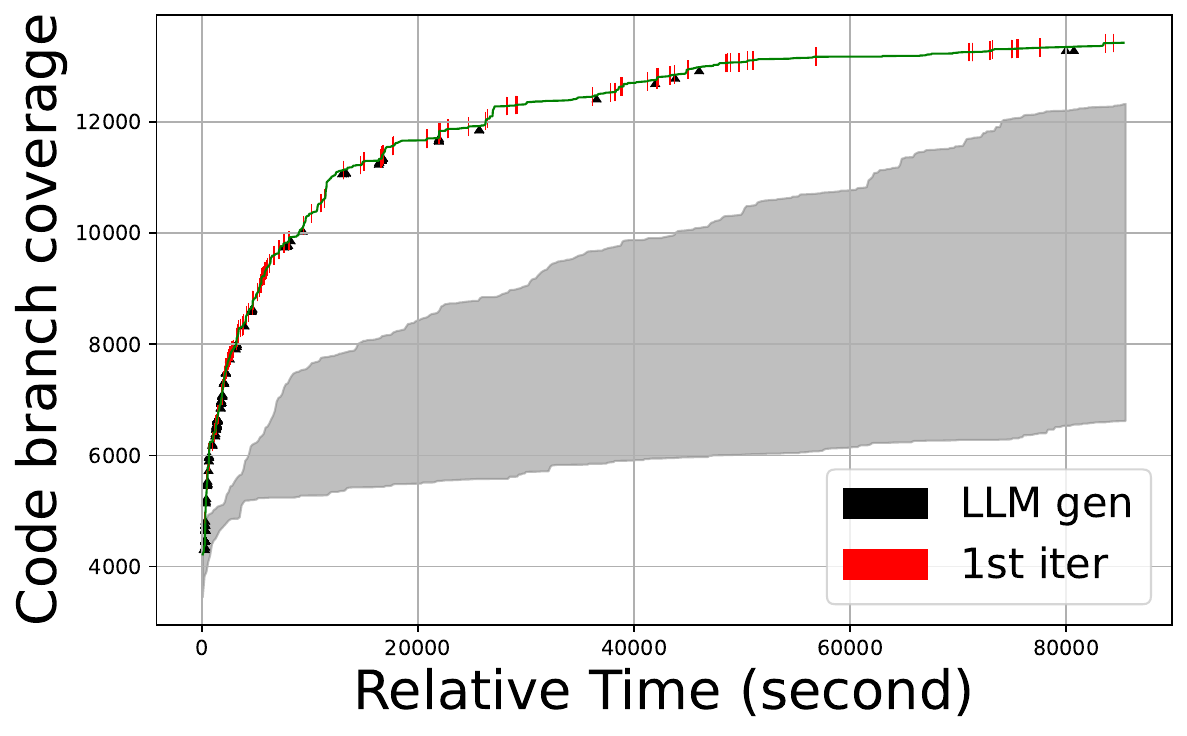}
\caption{Code coverage growth of {LLAMAFUZZ} and AFL++ on the binutils-nm target over time. Black triangles mark LLM-generated seeds, while the red vertical line highlights seeds directly sourced from LLM. The green line represents {LLAMAFUZZ}’s coverage, and the grey background indicates AFL++’s coverage range.}
\Description{Code coverage growth of {LLAMAFUZZ} and AFL++ on the binutils-nm target over time. Black triangles mark LLM-generated seeds, while the red vertical line highlights seeds directly sourced from LLM. The green line represents {LLAMAFUZZ}’s coverage, and the grey background indicates AFL++’s coverage range.}
\label{llmcoverage}
\vspace{-0.1cm}
\end{figure}

\begin{table}[h]
\centering
\scriptsize
\caption{Byte-level comparison of an original PNG seed and its AFL- and \textsc{LLAMAFUZZ}-mutated variants.}
\resizebox{\linewidth}{!}{%
  \begin{tabular}{p{0.11\linewidth} p{0.29\linewidth} p{0.29\linewidth} p{0.29\linewidth}}
    \toprule
    \textbf{Chunk} & \textbf{Original Bytes} & \textbf{AFL-Mutated Bytes} & \textbf{\textsc{LLAMAFUZZ}-Mutated Bytes} \\
    \midrule
    Signature &
    \texttt{89 50 4E 47 0D 0A 1A 0A} &
    \texttt{88 50 4F 47 0D 0A 1A 0A} &
    \texttt{89 50 4E 47 0D 0A 1A 0A} \\[2pt]

    IHDR &
    \texttt{00 00 00 0D 49 48 44 52 00 00 00 20 00} &
    \texttt{48 49 45 52 01 00 00 21 21 03 00 00} &
    \texttt{00 00 00 0D 49 48 44 52 00 00 00 5C 00} \\[2pt]

    Image\,data &
    \texttt{00 00 20 02 00 00 00 00 44 A4 8A C6 00 00 00 04 67 41 4D 41 00 00 9C 40 20 0D E4 CB 99 FF CC 3E 4C AF 15 00 63 64 5A 73 6D B6 62 3C 7A F1 A0 73 CD 40 20 B4 0A 0A 0A 0A} &
    \texttt{01 45 A5 8B C7 00 00 00 04 66 41 4C 41 00 00 9D 41 21 0C E5 CB 98 FF CD 3F 4D AE 14 00 62 64 5B 72 6C B7 63 3D 0B 0B 0B 0B} &
    \texttt{00 00 44 08 06 00 00 00 58 92 D5 E4 00 00 00 04 67 41 4D 41 00 00 B1 8F 14 83 66 05 54 05 05 05 05 56 DA 2E 30 01 01 95 05 56 DA 2E 30 01 01 95 4D 01 01 6A 98 01 01 08} \\
    \bottomrule
  \end{tabular}}
\label{tab:png-mutation}
\vspace{-0.1cm}
\end{table}

    \subsection{How did LLM augment the fuzzing process?}
    During the fuzzing process, seeds that can trigger new behavior are considered valuable and are used for further fuzzing. Therefore, understanding the relationship between these seeds and code coverage improvements is crucial for optimizing the fuzzing process. 
    \autoref{llmcoverage} shows the growth of coverage over time. 
    The seeds generated by LLM and the AFL++ seeds sourced from LLM seeds span the whole fuzzing process, which brings steady coverage improvement. This indicates that LLM-generated seeds not only directly impact the fuzzing process but also have a profound and indirect influence on its development. When compared to the grey area, which represents the interval of AFL++ coverage, {LLAMAFUZZ} achieved both higher and faster coverage. This observation further reinforces the LLM can understand the format structure and mutation strategies during fine-tuning instead of replaying the fine-tuning data. It also aligns with the outcomes from previous experiments conducted in~\autoref{exp2}.

    \subsection{Case study}
    To illustrate the advantages of LLMAFUZZ when handling complex data formats, we present an additional example in \autoref{tab:png-mutation} focusing on the highly structured binary format, PNG, and its target program, libpng. PNG files begin with an 8-byte signature and then a series of ordered, CRC-protected chunks (e.g., \texttt{IHDR}, \texttt{IDAT}, \texttt{IEND}).  Any effective mutation must preserve format validity while modifying semantic content to expose hidden bugs.

    \paragraph{\textbf{Maintains Structural Validity}} The PNG format's integrity hinges on its initial 8-byte fixed signature. As demonstrated in the comparison, AFL's bit-level mutation often corrupts this signature (changing '89' to '88'), rendering the file immediately invalid and preventing deeper program logic from being tested. In contrast, {LLAMAFUZZ} preserves the correct signature, ensuring the mutated input can be processed by the target program.
    
    \paragraph{\textbf{Semantically-Aware Chunk Mutation}} PNG files are composed of well-defined chunks with specific ordering and formatting requirements. {LLAMAFUZZ} selectively targeting important chunks, such as IHDR, and mutating their semantic content (altering the filter method from '20' to '5c') without disrupting the essential chunk boundaries. Conversely, AFL's random bit-level mutations frequently corrupt chunk lengths, types, or CRC checksums, leading to invalid inputs.

    By preserving structural correctness and injecting semantically meaningful mutations, {LLAMAFUZZ} generates inputs that explore the target program's codebase more deeply. This increases the probability of uncovering vulnerabilities that random mutation strategies would likely miss.

\section{Related Work}
\subsection{Fuzzing}
    
    White-box fuzzers utilize program analysis to improve the code coverage to explore certain code regions, which can be efficient in revealing vulnerabilities in complex logic. 
    WhisperFuzz~\cite{borkar2024whisperfuzz} introduces a static analysis method designed specifically to detect and locate timing vulnerabilities in processors. The tool focuses on evaluating the coverage of microarchitectural timing behaviors, providing a targeted and comprehensive assessment that aids in identifying potential security risks associated with timing flaws.
    
    However, program analysis and defining specialized seed generation grammar could be extremely time-consuming. Greybox fuzzer combines the effectiveness of white-box fuzzer and the efficiency of black-box fuzzer. It leverages instrumentation to get feedback from target programs and leading fuzzers to generate more valuable seeds resulting in higher code coverage. Greybox fuzzers usually combined with mutation strategies rely on iterative modifications of existing seeds to produce novel fuzzing inputs. 
    In addition to basic mutations, recent researchers have developed complex transformations to maintain type consistency~\cite{jain2018tiff,chaliasos2022finding}, adding historical bug-triggering code snippets~\cite{holler2012fuzzing,zhao2022history}, and coverage feedback~\cite{aschermann2019nautilus,fioraldi2020afl++} for improved testing efficiency. American Fuzzy Lop (AFL)~\cite{afl} and its variations~\cite{fioraldi2020afl++,lyu2019mopt,crump2023libafl}, employ genetic algorithms with a fitness function to prioritize fuzzing inputs for further mutations aimed at enhancing coverage, concentrating on byte-level changes.



\subsection{Coverage-guided greybox fuzzing}

    To overcome the inherent randomness challenges in fuzzing, researchers suggest using bit-map to record coverage information as feedback to more effectively guide the fuzzing process~\cite{afl}. Since vulnerabilities cannot be detected in uncovered paths, focusing on expanding the coverage of execution paths is a reasonable step toward improving the performance of fuzzing techniques.
    
    Given a program under test and a set of initial seeds, the coverage-guided greybox fuzzing process mainly consists of four stages. 
    
    \textbf{Seeds queue:} a seed was selected from the seeds pool for mutation.
    
    \textbf{Seed mutation:} the selected seed was mutated by various mutation strategies to generate new test seeds.
    
    \textbf{Execution:} Execute the current seed into the program.
    
    \textbf{Behavior monitoring:} Each new seed will be fed into the instrumented program for execution and evaluated by coverage metric. If the seed triggers a new coverage, it will be added to the seeds queue for further fuzzing.
    
    As the fuzzing loop continues, more code branches will be reached, which holds the potential to trigger a bug~\cite{wang2019sensitive}.

\subsection{Fuzzing for structured data}

    Coverage-guided greybox fuzzing has been effective in identifying vulnerabilities in many real-world programs. However, with the increasing complexity of software development, many programs use highly structured data in special formats, which poses significant challenges for traditional fuzzing techniques. Traditional fuzzers primarily perform mutations at the bit level, requiring excessive attempts to mutate such structured data effectively. Moreover, blind random mutation strategies often disrupt the consistency of data formats, leading to the generation of numerous inefficient and ineffective test cases.

    Fuzzers for structured data~\cite{wang2019superion, le2021saffron, mallissery2023demystify, zhang2024wasmcfuzz, meng2024large, koffi2024structuredfuzzer, shi2023aifore} can accurately identify the target input format. They can generate test cases that maintain the consistency of the format. This approach ensures that the generated test cases are not only valid but also effective in triggering and exploring potential vulnerabilities or issues within the application. Three grammar-aware mutation operators have been found to be particularly effective in uncovering deep bugs~\cite{srivastava2021gramatron,aschermann2019nautilus}: random mutation, which involves selecting a random non-leaf non-terminal node and creating a new context-free grammars derivation subtree. Random recursive unrolling, which finds recursive production rules and expands them up to $n$ times. Splicing, which combines two inputs while preserving their syntactic validity. In addition, Langfuzz~\cite{holler2012fuzzing} combines grammar-based fuzz testing and reusing project-specific issue-related fragments, maintaining the integrity of format and having a higher chance to cause new problems than random input.
    QuickFuzz~\cite{grieco2016quickfuzz} leverages Haskell's QuickCheck and the Hackage to fuzz structured data. This integration, combined with conventional bit-level mutational fuzzers, negates the need for an external set of input files and eliminates the requirement to develop specific models for the file types being tested. Alternatively, WEIZZ~\cite{fioraldi2020weizz} employs a chunk-based mutator to generate and mutate inputs for unknown binary formats. Nevertheless, WEIZZ struggles to handle grammar-based formats such as JSON, XML, and programming languages.

\subsection{Augment fuzzing through machine learning}
    Current research primarily concentrates on two aspects: employing machine-learning models as generators and leveraging machine-learning models to guide the fuzzing process.

    C. Pérez~\cite{perez2024compressed} explored the ability of Compressed-Language Models (CLMs) to interpret files compressed by standard file formats. Their findings revealed that CLMs are capable of understanding the semantics of compressed data directly from the byte streams, opening a new path for processing raw compressed files. In a related study, CHATFUZZ\cite{hu2023augmenting} investigates the mutation capabilities of LLM on text-based seeds, achieving 12.77\% edge coverage improvement over the SOTA greybox fuzzer (AFL++).
    Similarly, SmartSeed~\cite{lyu2018smartseed} combines deep learning models to generate new inputs for evaluating 12 different applications.
    
    Prior work~\cite{Eom2024CovRLFJ} integrates an LLM-based mutator with a reinforcement learning approach, utilizing the Term Frequency-Inverse Document Frequency technique to develop a weighted coverage map. This method capitalizes on coverage feedback to enhance the effectiveness of the mutation process. Similarly, Xia et al.~\cite{xia2024fuzz4all} introduce an auto-prompting phase that employs LLMs to produce and mutate test cases across six programming languages. Their findings indicate that LLMs can surpass the coverage achieved by cutting-edge tools.

    Additionally, WhiteFox~\cite{yang2023white} employs dual LLMs within their framework: one analyzes low-level optimization source code to inform optimization strategies, while the other generates test programs based on this analysis. CHATAFL~\cite{meng2024large} utilizes LLMs to understand protocol message types and assesses their ability to identify "states" in stateful protocol implementations. LLM4FUZZ~\cite{shou2024llm4fuzz} leverages LLMs to guide fuzzers towards more critical code areas and input sequences that are more likely to reveal vulnerabilities, showcasing the potential of LLMs in prioritizing and refining fuzzing efforts.

\section{Conclusion}

Mutation is a critical step in greybox fuzzing that directly impacts performance. Although randomized bit-level mutations work well in many cases, state-of-the-art fuzzers struggle with structured data because generating valid, highly structured inputs typically requires many attempts and relies heavily on randomness. In this paper, we propose leveraging LLMs to learn structured data patterns and guide seed mutation. We evaluate {LLAMAFUZZ} on a ground-truth fuzzing benchmark and a diverse set of real-world programs handling structured data. Our method achieves significantly higher coverage and identifies 47 unique bugs across all trials. These findings confirm LLAMAFUZZ’s effectiveness in structure-aware mutation.


\bibliographystyle{ACM-Reference-Format}
\bibliography{sample-base}


\end{document}